\documentclass[letter]{aa} % for the letters 
\usepackage{graphicx}
\usepackage{txfonts}
\begin{document}
   \title{The diffuse radio filament} 

   \subtitle{in the merging system ZwCl 2341.1+0000}

   \author{G. Giovannini \inst{1,2}, A. Bonafede \inst{1,2},
           L. Feretti \inst{2},
           F. Govoni \inst{3}, M. Murgia \inst{3}}

 \institute{(1) Dipartimento di Astronomia, via Ranzani 1, 40127 Bologna, I \\
            (2) Istituto di Radioastronomia-INAF, via P.Gobetti 101,
                40129 Bologna, I \\
            (3) Osservatorio Astronomico di Cagliari - INAF,
                Strada 54, Loc. Poggio dei Pini, 09012 Capoterra (Ca), I
             }

\authorrunning{Giovannini et al.}

   \date{ October 14, 2009 }

\abstract{In some clusters of galaxies, a diffuse non-thermal emission is 
present, not obviously associated with any individual galaxy. These sources
have been identified as relics, mini-halos, and halos according to their 
properties and position with respect to the cluster center. Moreover in a few
cases have been reported the existence of a diffuse radio emission not 
identified with a cluster, but with a large scale filamentary region.
}
{The aim of this work is to observe and discuss the diffuse radio emission
present in the complex merging structure of galaxies ZwCl 2341.1+0000.
}
{We have obtained VLA observations at 1.4 GHz to derive a deep radio image
of the diffuse emission.
}
{Low resolution VLA images show a diffuse radio emission associated to the 
complex merging region
with a largest size = 2.2 Mpc. In addition to the previously reported
peripheral radio emission, classified as a double relic, 
diffuse emission is detected along the optical filament of 
galaxies.
}
{The giant radio source discussed here shows that magnetic fields and 
relativistic particles are present also in filamentary structures.
Possible alternate scenarios are: a giant radio halo in between two
symmetric relics, or the merging of two clusters both hosting a central
radio halo. 
} 


   \maketitle
%
%________________________________________________________________

\section{Introduction}

Diffuse non-thermal radio sources with steep spectra have been found in a
relatively large number of clusters of galaxies (see e.g. \cite{gio09}
and references there in). These sources are not directly associated
with the activity of individual galaxies and are related to physical 
properties of the whole cluster. According to their properties and location
they are commonly classified as Relics and Halos.

Relics have been found at the cluster peripheries and the most common
interpretation is that they are related to the presence of shocks originated
by cluster mergers. Halos are located in the central cluster regions and
have always been 
found in clusters with evidence of merger activity, which plays 
an important role in particle re-acceleration providing the energy that powers
these sources (\cite{bru09}, \cite{fg08}).

In addition some evidence has been found of the existence of non thermal
emission on even larger scales. Bridges of radio emission have been observed 
in the regions between relics and halos in a few clusters, including Coma
(\cite{kim89}, \cite{gio90}), A2255 (\cite{fer97}), and 
A2744 (\cite{gov01}).
Diffuse emission have been found at large distance from A2255 (\cite{piz08}), 
and A2256 (\cite{wee09a}).

A convincing evidence of radio emission from a filament structure was reported
by \cite{bag02}, identified with the multi-Mpc scale filamentary
network of galaxies in the ZwCl2341.1+0000 region at z = 0.27. 
The authors discuss the presence of large scale shocks originating in the
accretion flows of intergalactic gas, inferred from the Mpc scale diffuse radio
emission.
\cite{wee09b} presented GMRT observations at 610 MHz, 241 MHz 
and 157 MHz of this 
region, combined with X-ray and optical data. The radio images show 
two diffuse 
sources to the north and south of the cluster position which they classified as
double radio relics. These relics are perpendicular to the X-ray axis which
can be considered the merger axis. They are suggested to be due to outward
travelling shocks caused by a major merger event. The distance between the two
relics is $\sim$ 2.2 Mpc. 

We present here new low resolution VLA images of the diffuse radio emission
at 1.4 GHz. 
Despite of the higher frequency, the better sensitivity to surface
brightness allowed us to detect very extended diffuse radio emission,
which  spans along the optical filament. 
We discuss its nature and properties.

The intrinsic parameters quoted in this paper are computed for
a $\Lambda$CDM cosmology with $H_0$ = 71 km s$^{-1}$Mpc$^{-1}$,
$\Omega_m$ = 0.27, and $\Omega_{\Lambda}$ = 0.73.
At z = 0.27 the angular conversion factor is 4.1 kpc/''.

\section{Radio images}
The ZwCl2341.1+0000 region was observed for 6 hrs with the VLA at 1.4
GHz in the D configuration on July 24 2008, at two different
pointings: RA=23h43m40s DEC=+00$^\circ$20'00'', and RA=23h43m50s
DEC=+00$^\circ$14'00'', to avoid high primary beam attenuation and
bandwidth smearing. Each field was observed for 3 hrs, and we moved the
pointing center every half an hour to obtain a better uv-coverage.
Calibration and imaging were performed with the NRAO Astronomical
Image Processing System (AIPS). 
The sources 3C48, 3C138 and
2316+040 were used as primary flux-density calibrator, absolute
reference for the electric vector polarization angle and phase
calibrator respectively. Several cycles of self-calibration were
applied in order to remove residual phase and gain variations.  Images
of the total intensity (Stokes I) as well as of the Stokes parameters
U and Q were produced for the two pointings separately following the
common procedures: Fourier-Transform, Clean, and Restore.  The images
resulting from the two pointings were finally combined together using
the AIPS task LTESS.
We
then derived images of the polarized intensity $P=\sqrt{(U^2+Q^2)}$,
corrected for the positive bias, and of the polarization angle
$\Psi=0.5 arctan(U/Q)$. Calibration errors are of the order of 5\%.

\begin{figure}
   \centering
   \includegraphics[width=8.0cm]{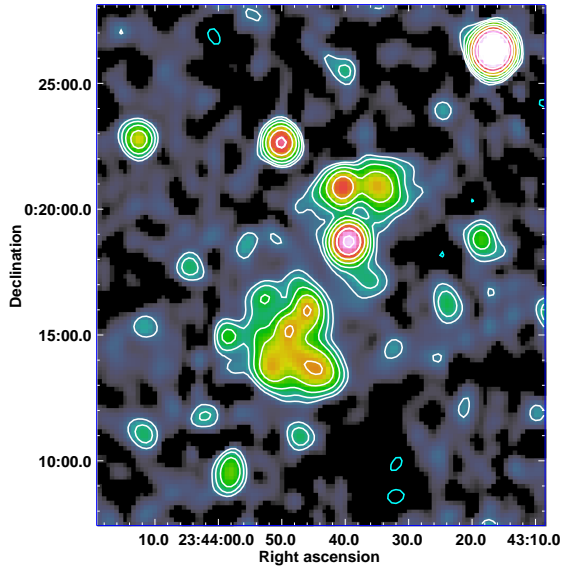}
      \caption{Radio emission from ZwCl 2341.1+0000 at 1.4 GHz
        observed at full resolution. The HPBW is
        50$''\times$43$''$. The
        first contour is drawn at 3$\sigma$ level (0.12 mJy/beam), 
other contours are
        spaced by a factor 2. The first negative contour (at 3$\sigma$ level)
is in cyan.}
         \label{fig:FR}
\end{figure}

  \begin{figure*}
   \centering
   \includegraphics[width=17.0cm]{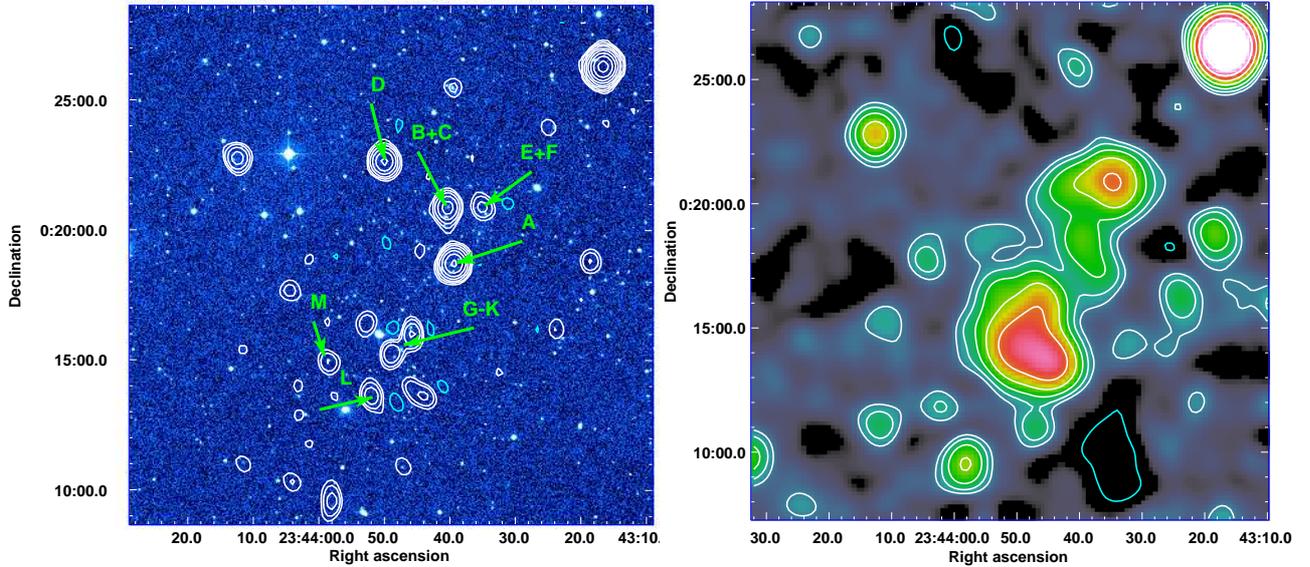}
         \caption{{\it Left:} The large scale galaxy distribution
           around ZwCl 2341.1+0000 is shown in colors. The optical image 
is from GSCII (\cite{las08}).
           Contours of the radio emission from discrete sources (see text) 
           are overlaid. The HPBW is 39$''
           \times$35$''$. The first contour is drawn at 3$\sigma$
           level (0.15 mJy/beam). Following contours are spaced by a
           factor of 2. The first negative contour at 3$\sigma$
level is displayed in
           cyan. Labels refer to the sources found by \cite{wee09b}. 
{\it Right:} Colors and contours refer to the
           radio emission after the subtraction of the discrete
           sources. The HPBW is 83$'' \times$75$''$. The first contour
           is drawn at 3$\sigma$ level (0.15 mJy/beam). Following
           contours are spaced by a factor of 2. }
         \label{fig:LR}
   \end{figure*}

\subsection{Total intensity radio emission}
In Fig. \ref{fig:FR} we show the final image obtained with natural weights.  
The
diffuse emission is easily visible even if a large number of discrete
sources are present.  To obtain the image of the diffuse source, and 
measure its flux, 
we subtracted unrelated discrete sources in the uv
plane.  To this aim, we produced high resolution images with uniform
weight (ROBUST = -5), after cutting baselines shorter than 2$K\lambda$. 
In these images only discrete sources are present, whereas
the extended diffuse emission is resolved out (see Fig. \ref{fig:LR}, 
left panel).

Discrete sources present in our images are in agreement with the list
of unrelated sources found by \cite{wee09b} using GMRT data
at lower frequencies, but at higher resolution. A comparison between
GMRT flux densities at 610 MHz and at 1.4 GHz, taking into account the
different angular resolution and uv-coverage, gives spectral indexes in the
range 0.8 - 1.5, expected from extra-galactic radio sources
(for the strongest subtracted source (A)  $\alpha_{0.6}^{1.4}$ = 1.2). 
Clean components were subtracted from the uv data.  With the new data
sets we produced low resolution images to enhance the low brightness
extended emission.  Final images were combined and corrected for the
primary beam attenuation. The final combined, and primary beam
corrected image is shown in the right panel of Fig. \ref{fig:LR}. 
In this image, the radio emission is continuous,  and the  gap between the 
northern and  southern regions visible in Fig. \ref{fig:FR} is no longer
present, clearly because of a positive noise in this area, 
enhanced by the slightly larger beam.

\subsection{Polarized emission}
\begin{figure}
 \centering
 \includegraphics[width=6.0cm]{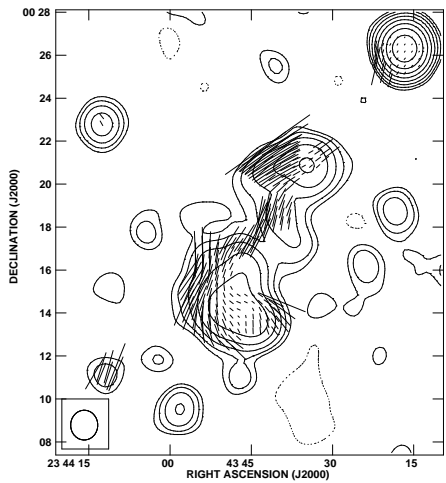}
    \caption{Polarized emission from ZwCl 2341.1+0000 at 1.4
      GHz. Contours show the total-intensity emission at low
      resolution (see Fig. \ref{fig:LR}. Lines refer to the E vectors
      Their orientation represent the projected E-field not corrected
      for the galactic rotation.  Their length is proportional to the
      fractional polarization: 1$''$ corresponds to 0.2\%.}
       \label{fig:pol}
 \end{figure}

We investigated the presence of polarized emission from the radio
diffuse emission. We produced Stokes Q and U low resolution
images. The discrete sources were subtracted as explained above for the
total intensity image. We derived the polarization angle image and the
polarization intensity image without imposing any cut. From the
polarization intensity image we derived the fractional polarization
image, and we considered as valid pixels those whose signal-to-noise
ratio was $>$3. This cut is done in order to get rid of possible
spurious polarization. The resulting image is shown in
Fig. \ref{fig:pol}. We can gather that there is a detection of
polarization in the northern and southern components, as well as  in the
central region. 

\section{Results}

In the final image, obtained  after the subtraction of discrete sources, 
an extended emission is detected, which is consistent
with the result of \cite{bag02}. Owing to the better
sensitivity of our image, the diffuse source is well defined. The radio 
morphology is elongated, clearly following the distribution of the optical
galaxies and of the X-ray emission, shown by \cite{wee09b}.
We note that the regions of highest brightness are coincident with 
the two relics presented by \cite{wee09b}.
In the central region, we detect a low surface brightness emission,
at the level of  0.4 -- 0.6 mJy/beam (corresponding to $\sim$ 
7 10$^{-5}$ mJy/arcsec$^2$), which could be detected at 610 MHz with a 
15'' beam only if the spectral index $\alpha ^{1.4}_{0.6}$ is 
steeper than 2.5 - 3.

The total size of the diffuse emission is $\sim$ 2.2
Mpc. The measured total flux at 1.4 Ghz is 28.5 mJy,
corresponding to a radio power log $P_{1.4}$ = 23.66 W/Hz. 
The radio emission is irregular and decreases from the two bright outer
regions toward the cluster center.
A plateau of radio emission, at the level of about 0.6 mJy/beam 
is present at the location of the southernmost X-ray peak, detected by
Chandra and published in Fig 1 of \cite{wee09b}. 

We detect polarized emission from large areas of the diffuse radio
source, both from the outer and the central regions.  The polarized
flux is more prominent in the eastern side of the extended source.
Once the discrete sources have been subtracted, the total polarized
flux is $\sim$2.4 mJy.  The polarized percentage in the northern and
in the southern bright regions is $\sim$15\% and 8\% respectively, while
the mean fractional polarization in the central region is $\sim$11\%.
The polarization vectors are very regular and oriented toward the
NE-SW direction in the northern source region.   In the other
regions they  follow the eastern edge of the total intensity
emission  still showing some level of ordering. 

From present images and following the detailed discussion by 
\cite{bag02}, and \cite{wee09b}, we ruled out the 
identification of this extended emission as a giant radio galaxy, and
we compared its properties with those
of radio halos presented in \cite{gio09}. The 2.2 Mpc size is 
remarkably high, but it is comparable to that of the radio halos in A2163 
and 1E0657-56. On the contrary, the total radio power at 1.4 GHz is 
considerably lower than the expected value for a giant radio halo from the 
correlation between halo's size and radio power shown
in \cite{gio09} (and references therein).

\section{Discussion}

Radio emission of very low brighthness is detected in the
ZwCl2341.1+0000 complex, owing to the high sensitivity to surface
brightness of the present observations.  The data reported in this
paper indicate the existence of diffuse radio emission along the whole
optical filament, as first reported by \cite{bag02}, in addition
to the bright radio regions located at the outer source boundary,
where \cite{wee09b} suggested the existence of two relics.
The low angular resolution of the present data does not allow us to
separate the the radio emission of these features from the rest of
the diffuse emission.

The most obvious interpretation is, as suggested by \cite{bag02},
that this region is witnessing the process of a large scale
structure formation, where cosmic shocks originated by a complex
multiple merger are able to accelerate particles and amplify seed
magnetic fields.  The large-scale radio emission would originate from
large-scale shocks and possibly turbulence, and the identification
of single radio source components  as relics, would be problematic.
The existence of significant polarized emission image can be easily 
explained in this scenario. The shocks connected to large scale
filaments are indeed expected to compress the magnetic field,
allowing the detection of polarized emission also at such low
resolution. Furthermore, we note that the lower gas density of
filaments with respect to clusters would cause less depolarization,
and internal Faraday rotation, that could explain the ordered features
revealed by the polarization image. The high fractional polarization
in the eastern side could be related to the direction and dynamics of the
merging process. 

An alternate possibility could be that the extended emission in this
cluster actually consists of two peripheral relics and a central radio
halo, as in the clusters A1758 (\cite{gio09}) and
RXCJ1314.4-2515 (\cite{fer05}, \cite{ven07}).  In this scenario,
shocks originated in the merging processes of this complex structure
could give origin to the peripheral relics, while a large scale
turbulence at the cluster center as enhanced by X-ray data (\cite{wee09b}) 
could give origin to the central halo.  Although it is
not obvious to separate in ZwCl12341.1+0000 the relic and halo
emission, we estimate that the radio halo size is $\sim$ 4' ($\sim$ 1
Mpc) with a flux density $\sim$ 10 mJy (log $P_{1.4}$ 23.2 W/Hz), and
the Northern and Southern relics have a flux density $\sim$ 5 mJy and
13 mJy, respectively. The spectral index of the two relics between 1.4
GHz and 610 MHz, estimated from the present data and those by 
\cite{wee09b}, is $\sim$ 1.2 (for both relics). This value is higher
with respect to $\alpha_{241}^{610}$ = 0.49 and 0.76 for the N and
S relics respectively reported by \cite{wee09b}. Although this may be due 
to a genuine flattening at low frequency, we note that because of different UV
coverages and sensitivities, it is not possible to derive firm conclusions 
about the spectral index properties.

From the image presented in this paper, we derive that the central
radio halo is likely to show a quite irregular structure, however its
radio size and power are in good agreement with the correlation
between radio power and size reported by \cite{gio09}.  Also
the two possible relics are peculiar, indeed their shape is not
regular and elongated as in typical relics like 1253+275 in the Coma
cluster (\cite{gio91}) or the double relics in A3376 (\cite{bag05}), and
A3667 (\cite{rot97}). 
However, exceptions are known in the literature as the
relics in A1664 (\cite{gov01}), and A548b (\cite{fer06}). 
The presence of polarized
emission in the two relics is in favour of this scenario, whereas the
detection of polarization in the central halo would be against it. Indeed,
radio halos are typically unpolarized, with upper limits of the
order of few percent. The only two cases known so far where polarized
emission has been detected from radio halos are the cluster Abell 2255
(\cite{gov05}) and MACS~J0717+3745 (\cite{bon09}).

As a third possiblity, we wish to remember the case of two halos
found for the first time in the double cluster system A399/A401
(\cite{mur09}). These two clusters are probably
in a pre-merging phase: the X-ray excess, the slight temperature
increase in the region between the two clusters, and the
relatively high metallicity of the hot IGM in the same region are
evidence of a physical link between this pair of clusters (\cite{mur09},
and references there in).  In this scenario, we may
interpret the complex radio emission in ZwCl2341.1+0000 as an evolved
case of two clusters, both hosting a radio halo.  During the ongoing
merger process, the two radio halos would appear as brighter diffuse
radio emitting regions, connected by a faint radio bridge, which could
orginate from the cluster interaction as in the bridge in the Coma
cluster (\cite{kim89}). After the merger, the two
clusters would eventually create a giant radio halo.  A major concern
about this scenario is that from the optical isodensity contours
presented by \cite{wee09b}, in the brightest radio emitting
regions there is no evidence of a galaxy overdensity which is the
expected signature of the two merging clusters. Furthermore, the
detected polarization flux could not obviously be reconciled with this
scenario.

\section{Conclusions}

Diffuse radio emission identified with the ZwCl2341.1+0000 complex
is detected along the whole filament of galaxies, in agreement with 
the results of \cite{bag02}.
We confirm that the regions of highest radio brightness are located at
the outer source boundary, where \cite{wee09b} suggested the
presence of two relics.  The angular resolution of the present data
does not allow us to separate the radio emission of the various
features.
We suggest three possible scenarios:

\begin{enumerate}

\item
The cluster is the site of cosmic shocks originated by the multiple
mergers during the large scale structure formation, as proposed by
\cite{bag02}.  In this case, the radio emission would be
related with the galaxy filament as a whole.
The polarized emission image could be easily reconciled with this
scenario, as the shocks connected to large scale filaments are
expected to compress the magnetic field to a high degree of ordering.

\item
The diffuse radio emission could consist of a central radio halo and
two opposite radio relics, as found in A1758 or
RXCJ1314.4-2515. Despite of the irregular structure of the central
halo, its size and power would be consistent with the correlation
between radio power and size reported by \cite{gio09}.  The
structure of the two relics is different from that of typical
elongated relics, however some other cases of irregular relics are
found in the literature (A1664, A548b).  The fractional polarization
detected in the diffuse source is consistent with the polarization
of relics, whereas the radio halos are generally unpolarized (but
see A2255 and MACS~J0717+3745).

\item
The complex structure could derive from a double cluster, hosting two
radio halos, similar to the cluster system A399/A401 (\cite{mur09}),
but at a later stage of merging, therefore the system has
developed a radio bridge between the two interacting clusters. The
presence of polarized flux cannot be easily reconciled with this
scenario, as radio halos are generally unpolarized.  Moreover, from
the optical isodensity contours presented by \cite{wee09b},
in the brightest radio emitting regions there is no evidence of a
galaxy overdensity which is the expected signature of the two merging
clusters.
\end{enumerate}

\begin{acknowledgements}
The National Radio Astronomy
Observatory is operated by Associated Universities, Inc., under cooperative
agreement with the National Science Foundation.
\end{acknowledgements}

\end{document}